\documentclass{aa}  
\def\ha{H$\alpha$}
\def\rha{$r$$-$H$\alpha$}
\def\ri{$r$$-$$i$}
\def\jh{$J$$-$$H$}
\def\hk{$H$$-$$K_S$}
\def\magcir{\ \raise-2.truept\hbox{\rlap{\hbox{$\sim$}}\raise5.truept
    \hbox{$>$}\ }}
\usepackage{graphicx}
\usepackage{txfonts}
\begin{document}
%
%\title{Symbiotic stars from IPHAS}
\title{IPHAS and the symbiotic stars}
\subtitle{I. Selection method and first discoveries\thanks{Based on
observations obtained at the 2.5m~INT telescope of the Isaac Newton
Group of Telescopes in the Spanish Observatorio del Roque de Los
Muchachos of the Instituto de Astrof\'\i sica de Canarias.
This publication makes use of data products from the Two Micron All
Sky Survey, which is a joint project of the University of
Massachusetts and the Infrared Processing and Analysis
Center/California Institute of Technology, funded by the National
Aeronautics and Space Administration and the National Science
Foundation. This research has also made use of the SIMBAD database,
operated at CDS, Strasbourg, France.}}

\author{R.L.M. Corradi\inst{1,2}
          \and
        E.R. Rodr\'\i guez--Flores\inst{2,3}
          \and
        A. Mampaso\inst{2}
          \and
        R. Greimel\inst{1}
         \and
        K. Viironen\inst{2}
	\and\\
	J.E. Drew\inst{4,5}
	\and
	D.J. Lennon\inst{1,2}
	\and 
	J. Mikolajewska\inst{6}
         \and
        L. Sabin\inst{2}
         \and
        J.L. Sokoloski\inst{7}
	}

   \offprints{R. Corradi}

   \institute{Isaac Newton Group. P.O. Ap.\ de Correos 321,
         38700 Sta. Cruz de la Palma, Spain\\ 
   \email{rcorradi@ing.iac.es}
   \and
        Instituto de Astrof{\'{\i}}sica de Canarias, 38200 La Laguna, 
        Tenerife, Spain
   \and
        Instituto de Geof\'\i sica y Astronom\'\i a, Calle 212, N. 2906, 
        CP 11600, La Habana, Cuba
   \and
        Imperial College of Science, Technology and Medicine, Blackett 
        Laboratory, Exhibition Road, London, SW7 2AZ, UK 
   \and
        Centre for Astrophysics Research, STRI, University of Hertfordshire,
        College Lane, Hatfield, AL10 9AB, UK
   \and
       N. Copernicus Astronomical Center, Bartycka 18, 00-716 Warsaw, Poland
   \and
       Columbia Astrophysics Laboratory, USA
             }

\date{Received 5 November 2007 / Accepted 10 December 2007}

\abstract{The study of symbiotic stars is essential to understand important
aspects of stellar evolution in 
interacting binaries. Their \protect{\it observed} population in the Galaxy is
however poorly known, and is one to three orders of magnitudes smaller
than the \protect{\it predicted} population size.}
{IPHAS, the INT Photometric \ha\ survey of the Northern Galactic
plane, gives us the opportunity to make a systematic, complete search for 
symbiotic stars in a magnitude-limited volume, and discover a 
significant number of new systems.}
{A method of selecting candidate symbiotic stars by
combining IPHAS and near-IR (2MASS) colours is presented. It allows us to
distinguish symbiotic binaries from normal stars and most of the other
types of \ha\ emission line stars in the Galaxy. The only exception are 
T~Tauri stars, which can however be recognized because of their
concentration in star forming regions.}
{Using these selection criteria, we discuss the classification of a
list of 4338 IPHAS stars with \ha\ in emission. 1500 to 2000 of
them are likely to be Be stars. Among the remaining objects, 1183
fulfill our photometric constraints to be considered candidate
symbiotic stars. The spectroscopic confirmation of three of these
objects, which are the first new symbiotic stars discovered by IPHAS,
proves the potential of the survey and selection method.}  {}


\titlerunning{IPHAS and the symbiotic stars. I.}
\authorrunning{R.L.M. Corradi et al.}
\maketitle

%________________________________________________________________

\section{Introduction}

Symbiotic stars are the interacting binaries with the longest orbital
periods. They are composed of a compact star, in most cases a hot
white dwarf, accreting from the wind of a cool giant companion.  Part
of the giant's wind is ionized by the white dwarf, producing the
composite spectrum containing both absorption features from a cool
stellar photosphere and emission lines from highly excited ions:  these
characteristics originally caused these objects to be named ``symbiotic''.
Depending on their near-IR colours, symbiotic stars are divided into
`stellar' (S) types, if colours are typical of red giant branch (RGB)
stars, or `dusty' (D) types, if their near-IR emission shows a
significant contribution from the warm dust known to be typical of evolved
asymptotic giant branch (AGB) stars. S-types account for around 80\% 
of the sample of known symbiotic stars.  The presence of an RGB or an AGB
star also determines the orbital separation at which the symbiotic
phenomenon occurs; orbital periods for the S-types are between 200 and
6000 days, while they are longer for the D-types ($>$20~yr, but
no single robust determination exists so far).  For more details on
the properties of symbiotic stars, see e.g. \cite{cmt03}.

A variety of phenomena occur in symbiotic stars that are relevant to
a number of important astrophysical problems. For example, symbiotic
stars have been proposed as potential supernova Ia progenitors (Munari
\& Renzini 1992, Hachisu et al. 1999). Also, symbiotic stars are
excellent laboratories for studying (i) thermonuclear outbursts
(nova-like accretion instabilities) under a wide range of conditions
(cf. Munari 1997); (ii) the powering mechanism of supersoft X-ray sources
(cf. Jordan et al. 1996); (iii) the collimation of stellar winds and the
formation of jets (cf. Tomov 2003) and (iv) bipolar (planetary) nebulae
(cf. Corradi 2003).

A crucial figure in the discussion of some of these topics is the total
number of symbiotic stars in the Galaxy.  This is basically
unknown. In fact, no systematic search for symbiotic stars in the
Milky Way has been done so far, and the present sample of 173 known
Galactic systems (\cite{b00}), plus another 26 suspected ones, is
mainly the result of occasional discoveries during the study of
peculiar, variable or erupting stars, or of sparse objective prism
surveys. This figure should be compared with the {\it predicted} total
number of symbiotic stars in the Galaxy, which spans two orders of
magnitude: 3$\times$$10^3$ (\cite{a84}), 3$\times$$10^4$ (\cite{k93}),
3$\times$$10^5$ (\cite{mr92}), and 4$\times$$10^5$ (\cite{mcm03}).

IPHAS, the INT Photometric \ha\ survey of the Northern Galactic plane
(\cite{d05}), gives us the opportunity to improve the
determination of this basic number.  The search for symbiotic systems
in IPHAS takes advantage of the generally strong \ha\ emission that
characterizes this class of objects.  In this paper, we define our
criteria for the detection of symbiotic stars using the
data from the IPHAS project, complemented by near-IR colours from the
2MASS survey.
The effectiveness of the method is illustrated by the spectroscopic
confirmation of three new symbiotic stars.

\section{The data}

\subsection{IPHAS}
\label{S-IPHASdata}

IPHAS is an international collaboration, whose aim is to produce a
complete, fully photometric, and spatially detailed \ha\ map of
the part of the Galactic Plane between latitudes $-5^\circ$ and
$+5^\circ$
that is visible from the Northern hemisphere.  The IPHAS observations
are obtained using the Wide Field Camera (WFC) at the prime focus of the
2.5m Isaac Newton Telescope (INT) on La Palma, Spain.  The WFC
consists of a mosaic of four 2k$\times$4k EEV CCDs, providing a field
of view of 34$\times$34 arcmin$^2$ with a sampling of 0$''$.33 per
pixel.  The IPHAS images are taken through three filters: a
narrow-band \ha\ ($\lambda_c$ = 6568{\AA}; FWHM = 95{\AA}) and two
broad-band Sloan $r$ and $i$ filters, with matched 120, 30, and 10 s
exposures, respectively. In this way, the magnitude range $13\leq r
\leq 20$ is covered for point sources (the fainter end at 10$\sigma$).
Pipeline data reduction and data distribution are handled by the
Cambridge Astronomical Survey Unit.
% (CASU, see {\it http://archive.ast.cam.ac.uk/}).
The presentation of the survey and further details can be found in
Drew et al. (2005).

At the time of writing this article, more than 90\%\ of the $\sim$1800
square degrees of the northern Galactic Plane to be covered by IPHAS
has been observed. A first photometric catalogue, containing more than
200 million objects, is about to be released (\cite{GS08}). From it, a
list of 4853 \ha\ emitting stars with $r<19.5$~mag has been extracted
by Witham et al. (2008). This is the sample that we consider in this
paper for our first search for symbiotic stars within IPHAS.
Independently, we are employing other methods to select both
point-like and extended \ha\ emitters from the IPHAS observations (see
e.g. Viironen et al. 2008): we will consider these additional samples
in following papers (note that most symbiotics have a stellar profile,
but a small fraction of them might be extended in the \ha\ filter due
to their resolved nebulosity, see Corradi 2003).

\subsection{Reference samples of known objects}

Unsaturated IPHAS data are available for only four known symbiotic
stars from the general catalogue of Belczy\'nski et al. (2000). They
are DQ~Ser, V352~Aql, V1413~Aql, and Ap~3-1.
%plus K~3--12, which is claimed to be a symbiotic star by \cite{as90}
%but is not included in the latest catalogue of \cite{b00}. ACCORDING TO
% JOANNA, ACKER SPECTRUM NOT GOOD ENOUGH TO UNDERSTAND.
In order to build a meaningful reference sample of known symbiotic
stars to be used to define our selection method, additional
observations were obtained at the INT using the same instrumental
setup as for IPHAS. Photometric data of 18 known symbiotic stars were
taken on June 2, 2004, and April 20, and September 17 and 18,
2005. Exposure times were tuned so as to avoid saturation of the
(generally bright) targets.  Reduction was done with IRAF.
Moreover, existing flux-calibrated spectra of 18 symbiotic stars (one
of which is in common with the sample above), obtained at different
epochs, were convolved with the WFC filters and instrument response curves
to derive \ha, $r$ and $i$ magnitudes in the IPHAS
photometric system. Where the $i$ magnitude cannot be derived from the
spectra, we adopted the \ri\ colour from \cite{m92}.
All together, we collected a sample of 39 known symbiotic stars (29 of
the S type and 10 of the D type) with magnitudes and colours fully
consistent with those produced by the photometric catalogue of IPHAS.
This constitutes the primary reference sample to develop our selection
method.

We have also investigated other classes of stars and nebulae which can
potentially be confused with symbiotic stars according to their IPHAS
colours, because of also having \ha\ in emission, or because their
spectral type is similar to that of the cool component of symbiotic
stars. Using IPHAS and additional observations, we have derived \ha,
$r$ and $i$ magnitudes for 67 planetary nebulae (PNe, Viironen et
al. 2008), 79 cataclysmic variables (Witham et al. 2006), and 518 Mira
variables. As Be stars are a frequent class of \ha\ emitting bright
stars in the Galactic plane (cf. Sect.~\ref{S-bestars}), they deserve
special attention: 18 of these stars in the low-reddening open
clusters NGC~663, NGC~869 and NGC~884, as well as 22 objects in the
more reddened cluster NGC~7419, turned out to have good IPHAS
photometric data and were adopted as the comparison samples for this
class of stars. Another frequent class of \ha\ emitters in star-forming 
regions are T~Tauri stars; their characteristics derived from IPHAS data
on Cyg OB2 are being investigated by Vink et al. (2008).  Finally, \ha\
emission is also observed in late K to M dwarfs with enhanced
chromospheric and coronal activity (dMe stars). However, their \ha\
emission is generally faint, with line equivalent 
widths\footnote{According to the standard definition, the equivalent
width of an emission line would be negative. However, in order to
simplify the discussion, in this paper we use a positive sign, 
so that a larger \ha\ equivalent width
corresponds to a stronger \ha\ emission, and to a larger \rha\
colour.} of a few \AA\ (\cite{hgr96}). These are much smaller than in
symbiotic stars (see Sect.~\ref{S-selectsymbio}), and for this reason
dMe stars are not further considered in this paper.

\subsection{Near-IR 2MASS magnitudes}
\label{S-2MASSdata}

As symbiotic stars contain a luminous cool giant, near-IR magnitudes
provide information both on their nature and support making distinctions 
between them and
other classes of objects. The division in the two main groups of S and
D types was indeed originally made using the $J$$-$$H$ and $H$$-$$K$
colours (Allen \& Glass 1974). Recently, the position of symbiotic
stars in the near-IR colour-colour diagram was discussed by
\cite{r06} and \cite{p07}.

We have therefore extracted from the Two Micron All Sky Survey (2MASS)
database the $J$, $H$ and $K_S$ magnitudes of 187 known symbiotic
stars, 288 PNe (most of them from \cite{rp05}), 95
cataclysmic variables, 1230 Be stars (1148 of which are from 
\cite{z05}), 487 T~Tauri stars (Dahm \& Simon 2005), and 121 Mira
variables (Rodr\'\i guez-Flores 2006).

Similarly, we have searched for 2MASS counterparts of the 4853 \ha\
emitters in \cite{w08}. For 4330 of them, we found a 2MASS source
within 1~arcsec from the IPHAS coordinates (note that the two sets of
data are calibrated into the same astrometric system). Also, the 2MASS
counterpart was identified for another 8 objects with slightly
worse astrometric match. This makes a total sample of 4338 \ha\
emitters with 2MASS magnitudes.

It should be noted that the detection limits of IPHAS
($r$$\le$19.5~mag for the objects in Witham et al. 2008) and of the
2MASS point-source catalogue ($K_S$$\sim$15~mag) roughly match the
characteristic colours of symbiotic stars.
We have considered a list of 71 known and bright symbiotic stars
($r$$\le$14.5~mag): 70\%\ of them have indeed an {\it observed}
optical-to-near-IR colour $(r-K_S)$$\ge$4.5. On the average, they are
presumably closer and less reddened than the new symbiotics stars that
we aim to detect with IPHAS, for which we then expect to be able to
find a 2MASS counterpart. We are therefore confident that adding 2MASS
data does not affect the completeness of our search for symbiotic
stars within IPHAS.

\begin{figure*}[!ht]
\centering
\includegraphics[width=16cm]{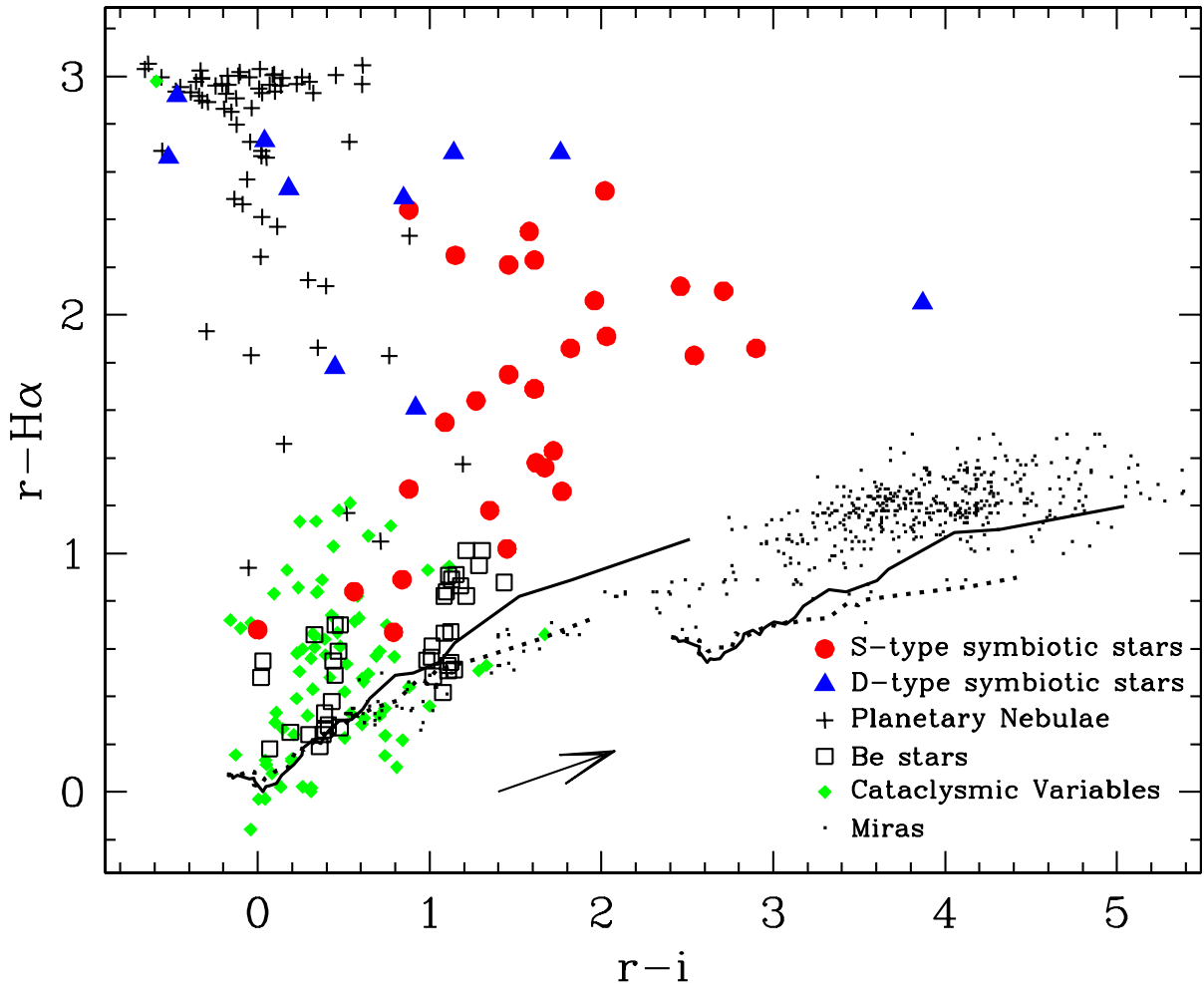}
\caption{IPHAS colour-colour diagram for different classes of
objects. The locus of main-sequence and RGB stars is indicated by the
solid and dotted lines, respectively.  Two sequences are shown,
corresponding to reddening values E(B-V)$=$$0$ (left) and $4$ (right),
respectively. The arrow indicates the reddening vector for normal stars:  
its length corresponds to 3~mag extinction in V.}
\label{F-IPHAScc}
\end{figure*}

\subsection{Follow-up spectroscopy}

As the present study was progressing, we started a campaign of
spectroscopic follow-up of the \ha\ emitters detected by IPHAS.
Accordingly a dozen candidate symbiotic stars, selected as
described in the next sections, were observed at the INT using the IDS
spectrograph, on nights of May 11, June 14, and September 9, 2006.
Grating R300V was used, which gives a reciprocal dispersion of
1.87~\AA\ per pixel of the 2kx4k EEV detector, and a spectral coverage
from 3800 to 8500~\AA\ (these figures slightly vary from night to
night).  The slit width was 1.1~arcsec projected on the sky, providing
a spectral resolution of 5~\AA. Exposure times were 30~min for the two
brighter sources discussed in Sect.~\ref{S-spectra}, and 2 hours for
the faintest and more reddened one. Several spectrophotometric
standards were observed during the night for relative flux
calibration.  Reduction was performed using the package {\it onedspec} in
IRAF. Note that the EEV CCD suffers from significant fringing redward
of $\sim$7000~\AA, which was not possible to remove with the
calibration frames obtained during those nights. Also, the flux
calibration is somewhat uncertain above 8000~\AA\ because of
significant optical aberrations at the edge of the large format CCD
used with IDS.

\section{Interpreting the IPHAS and 2MASS colour-colour diagrams}

\begin{figure*}
\centering
\includegraphics[width=16cm]{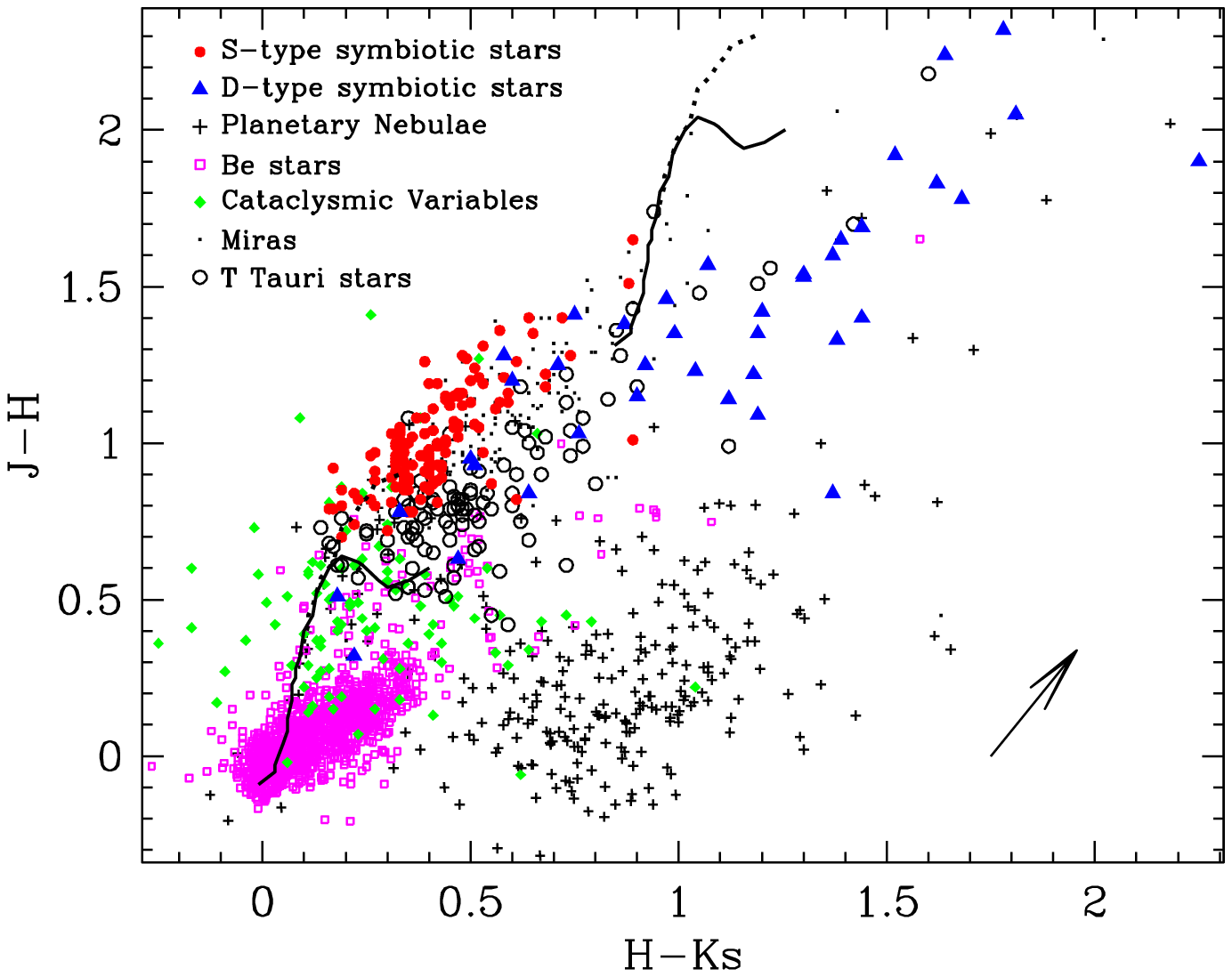}
\caption{2MASS colour-colour diagram for different classes of
objects. Symbols are the same as in Fig.~\ref{F-IPHAScc}, albeit
smaller owing to the larger number of objects. In addition, T~Tauri
stars are indicated by empty circles. Like in
Fig.~\ref{F-IPHAScc}, the locus of main-sequence and RGB stars is
indicated by the solid and dotted lines, respectively, for reddening
values E(B-V)$=$$0$ (lower-left sequence) and $4$ (upper-right
sequence). The arrow shows the reddening vector for normal stars
corresponding to 3~mag extinction in V.}
\label{F-2MASScc}
\end{figure*}

\subsection{The IPHAS colour-colour diagram}
\label{S-IPHAScc}

Our primary tool for the selection of candidate symbiotic stars in the
Galactic Plane is the IPHAS \rha\ vs. \ri\ colour-colour diagram,
which is presented and thoroughly discussed in Drew et al. (2005). It
is shown in Fig.~\ref{F-IPHAScc} for the various classes of objects
considered.  In the diagram, the \rha\ axis mainly indicates
increasing values of the \ha\ emission line equivalent width, while
\ri, for normal stars, is essentially a sequence of increasing
spectral types and/or reddening.  In the figure, the locus of
main-sequence and RGB stars is indicated by the solid and dotted
lines, respectively (\cite{d05}).  Two sequences are shown,
corresponding to reddening values E(B-V)$=$$0$ (left) and $4$ (right),
respectively. The reddening vector for normal stars, adopted from
Howarth (1983) in the same way as by Drew et al. (2005), is indicated 
by the arrow: its length corresponds to 3~mag extinction in V.  We
note that, as shown by Drew et al. (2005),  at the higher extinctions 
probed by IPHAS ($6 < A_V < 12$, roughly) this vector flattens 
progressively and in a mildly spectral-type dependent manner.

The positions of the different classes of object are indicated as
follows. Symbiotic stars of the S type are shown as filled circles,
and the D types as triangles. In the case that several measurements of
the same object obtained at different epochs are available, we plot
the mean colours as the dispersion of symbiotic stars in the graph is
anyway large enough to cover their intrinsic variability.  PNe are 
indicated by crosses; cataclysmic variables by filled diamonds;
Be stars by empty squares; Mira variables by dots. In all
cases, we show the {\it observed} colours.
% JED - the following doesn't appear relevant to me
% as no reliable estimate of
% the reddening is available for the majority of objects.

As expected, given their emission-line spectra are superposed on
relatively weak continua, PNe stand out in the graph for their extreme
\rha\ colours. Most of them lie close to the \rha$\sim$3.1 limit which
is expected for pure \ha\ emission line stars (``ideal'' sources for
which all the flux in the \ha\ and $r$ bands comes from the \ha\
line). With smaller \ha\ excesses than PNe (with some overlap), but
still above the other classes of \ha\ emitters and normal stars, lie
the symbiotic stars. They span a significant range in \ri\ colour.  Be
stars and cataclysmic variables are located closer to the locus of
main-sequence stars with some moderate \ha\ excess (except for Nova
Aql 1995, which is now in the nebular phase and shows a large \rha\
colour). Mira variables (which only sometimes show the \ha\ line in
emission) are mainly found at the right of the diagram, owing to their
cool photosphere and frequent large reddening by circumstellar dust.
The class of T~Tauri stars is not shown in the figure, as they are
spread all over the diagram above the sequence of normal stars
(e.g. Vink et al. 2008). They show a broad range of spectral types as
well as \ha\ equivalent widths. The latter are in most cases smaller
than 500~\AA\ (see Fig.~5 in Dahm \& Simon 2005), which is lower than
for the most extreme symbiotic stars. Our strategy to separate them,
at least in a statistical sense, from the evolved stars on which we
are focusing our attention in this work, will be presented in
Sect.~\ref{S-clumping}.

The symbiotic stars with the smallest \rha\ colours
%(at the bottom left of the group of symbiotic stars) 
belong to the rarer subgroup of the {\it yellow} symbiotics,
containing G-K giants. Examples are BD-213873, AG~Dra and LT~Del.
% The other two in the lower region (blow the 0.7 inclined line) are
% Draco C-1 (carbon star) and SS 324 (=ALS(88) 2), M2 in Belczinski...
%This is because these stars are bluer than M giants (thus the smaller
%\ri\ colour), and with a more significant continuum flux in the $r$
%band (thus the smaller \rha).
Also, the figure suggests that the D-types have \rha\ colours
generally larger than the S-types, and more similar to PNe. This was
already remarked upon in the past (cf. Kenyon 1986), and in fact the
distinction between PNe and some D-type symbiotic stars is tricky
(cf. Corradi 2003). However, our present sample is limited to ten
objects.

Further information on the location of symbiotic stars in the IPHAS
diagram, albeit limited to the \ri\ colour, can be obtained from the
literature. \cite{m92} published R and I magnitudes for a
large sample of symbiotic stars. We have transformed these data to the 
Sloan system for the INT+WFC, and in this case we are able to correct for 
interstellar extinction using the data in Whitelock \& Munari (1992).  The 
resultant range of intrinsic \ri\ colours spanned by 56 symbiotic stars of 
S type is between $-0.1$ and $+2.8$~mag, while the range spanned by 
19 D-type systems is even larger, from $-0.7$ to $+3.6$~mag.

\subsection{The 2MASS colour-colour diagram}
\label{S-2MASScc}

The previous section has shown the potential of the IPHAS
colour-colour diagram to separate symbiotic stars from the vast
majority of stars. However, some overlap is present with classes of
\ha\ sources whose population in the Galactic Plane is much larger
than that of symbiotic stars. In addition, mixing between classes is
inevitably raised in the presence of photometric errors.

The 2MASS \jh\ vs. \hk\ colour-colour diagram provides an additional
resource to refine the selection of symbiotic stars (Rodr\'\i
guez-Flores 2006).  The various classes of objects, as well as normal
main-sequence and giant stars (Bessel \& Brett 1988), are indicated in
Fig.~\ref{F-2MASScc} with the same symbols as in Fig.~\ref{F-IPHAScc}.
In addition, we plot here as empty circles 104 classical T~Tauri stars
from \cite{ds05} with an \ha\ equivalent width larger than 40~\AA\
(roughly corresponding to the smallest \ha\ equivalent width shown by
the known symbiotic stars in Sect.~\ref{S-IPHAScc}).  All these
T~Tauri stars belong to the young cluster NGC~2264, which has a low
foreground extinction ($A_V=0.22$~mag), and is rich in pre-main
sequence \ha\ emitters (Dahm \& Simon 2005).  For all objects we plot
the {\it observed} colours.  As for the IPHAS diagram, the reddening
vector is computed from Howarth (1983) for the 2MASS filters J, H, and
K$_s$: the adopted shifts for a reddening of 1~mag in V are
0.113 and 0.069 for the \jh\ and \hk\ colours, respectively.

The bulk of PNe occupy a distinct locus at the bottom of the near-IR
diagram (see also \cite{rp05}), to the right of normal stars. Most of 
the cataclysmic variables and Be stars are instead clumped closer to the
locus of main-sequence stars. In particular, note the well-defined
sequence at the bottom of the diagram displayed by Be stars, which
extends to the right side of the unreddened main-sequence stars. We
will come back to this important feature in Sect.~\ref{S-bestars}.

\begin{figure}[!ht]
\centering
\includegraphics[width=8.8cm]{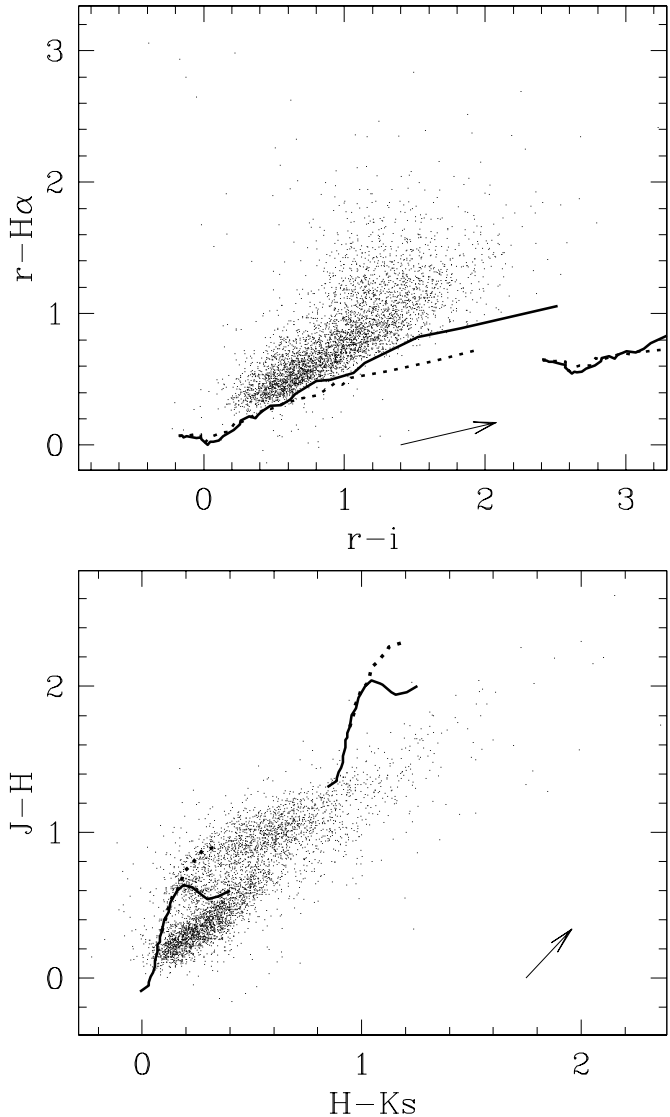}
\caption{The 4338 \ha\ emitters by \cite{w08} 
in the IPHAS (top) and 2MASS (bottom) colour-colour diagrams. The
loci of main-sequence and giant stars and the reddening vectors are
also indicated (see Fig.~\ref{F-IPHAScc} for a detailed explanation).}
\label{F-allwith}
\end{figure}

S-type symbiotic stars, according to their original definition (Allen
\& Glass 1974), have colours typical of RGB stars, with an upper-right
tail due to reddening. They form a compact cluster in the diagram,
except for few objects. The most extreme case is UV Aur (a carbon
star), which is displaced to the right of the S-types sequence, toward
the broad region which is occupied by T~Tauri stars, D-type symbiotic
stars, and some extreme Mira variables. The location of D-type
symbiotics is modeled by \cite{p07} with a variable combination of
stellar and warm dust continuum emission, plus some extinction.
% AG PEG REMOVED BECAUSE UNCERTAIN MAGNITUDE IN 2MASS
% AS221 REMOVED BECAUSE ITS CLASSIFICATION AS S-TYPE IS UNCERTAIN, THERE IS 
% DISAGREEMENT BETWEEN THE DENIS AND 2MASS MAGNITUDES AND JOANNA CONFIRMS IS 
% UNCERTAIN
Three objects, AS~201, V471~Per, and StHA~190, have significantly
smaller \jh\ colours than the bulk of symbiotics. Again, these are
yellow symbiotics, which tend to occupy the bottom part of the locus of 
symbiotic stars, as in the IPHAS diagram.
% THE D' TYPE SYMBIOTICS, DUSTY BUT WITH A G SPECTRAL
%TYPE GIANT.  THEY ARE AS 201, V471 PER, STHA_190, THE FIRST ONE IS THE
%LOWEST IN THE 2MASS DIAGRAM.  ALSO THE YELLOW OF THE S-TYPE ARE
%GENERALLY LOWER, E.G. AG PEG IS THE BLUE LOW POINT DETACHED ON THE
%RIGHT OF THE BULK OF S-TYPES
% ALSO, SOME S TYPE SEEM TO BE TOWARD THE D TYPE. 0.9 1.0 IS UV AUR,
%A CARBON STAR BUT OF S TYPE, DETACHED ON THE LOW RIGHT OF THE S-TYPES,
%AND AS 221, DETACHED ON THE UPPER RIGHT (CHECK LITERATURE, KENYON SYAS
%IT IS A S-TYPE OF SPECTRAL TYPE M4), MAYBE A D-TYPE? OR CHECK 2MASS
%FINDING CHARTS.

Some PNe fall in the region of D-type symbiotic stars. 
%They are He~2-25, M2-9, Mz~3, PM~1-339, Hen~2-171, Vo~1, H~1-12 and H~2-43. 
These are examples of possible misclassified PNe, which might
instead be symbiotic stars with extended nebulae as frequently
suggested in the literature (\cite{c95}, \cite{sk01}, \cite{scm07}).

Finally, T~Tauri stars form a broad sequence running parallel to, and on
right side of, S-type symbiotics, that extends well into the locus of the
D-type symbiotic stars. Note that T~Tauri stars with smaller
equivalent widths than those displayed, including the so-called
weak-line T~Tauri stars, would be clumped on the left-bottom end of
this sequence (Dahm \& Simon 2005).  The significant overlap with
symbiotic stars in both the 2MASS and IPHAS diagrams, makes T~Tauri
stars the most frequent ``contaminants'' in our search for symbiotics
stars. 

\section{Application to the list of IPHAS \ha\ emitters by \cite{w08}}
\label{S-application_to_witham}

The analysis of the IPHAS and 2MASS colour-colour diagrams presented
in the previous section allows us to discuss the nature of the 4338
IPHAS \ha\ emitters by \cite{w08} with 2MASS counterparts
(Sections~\ref{S-IPHASdata} and \ref{S-2MASSdata}). Their location in
the colour-colour diagrams is shown in Fig.~\ref{F-allwith}.  
Looking at the 2MASS diagram, there are some striking similarities with
the plot of known objects in Fig.~\ref{F-2MASScc}.
First, the 2MASS diagram of Fig.~\ref{F-allwith} shows a main
concentration of sources at its bottom, to the right of the unreddened
main-sequence stars. They form a dense band with a slightly lower
inclination than our adopted reddening vector. This is obviously similar to
the sequence defined by known Be stars in Fig.~\ref{F-2MASScc}.
Second, another inclined sequence of objects can be seen in
Fig.~\ref{F-allwith} at higher \jh\ values, in the region where
T~Tauri stars (frequent in the Galactic Plane) and symbiotic stars
(presumably less numerous) should be located.  

We therefore start our discussion by considering what appears to be
one of the most represented classes of objects in the list of
\cite{w08}: the Be stars.

\subsection{Be stars}
\label{S-bestars}

\begin{figure*}[!ht]
%\centering
\sidecaption
\includegraphics[width=6.5cm]{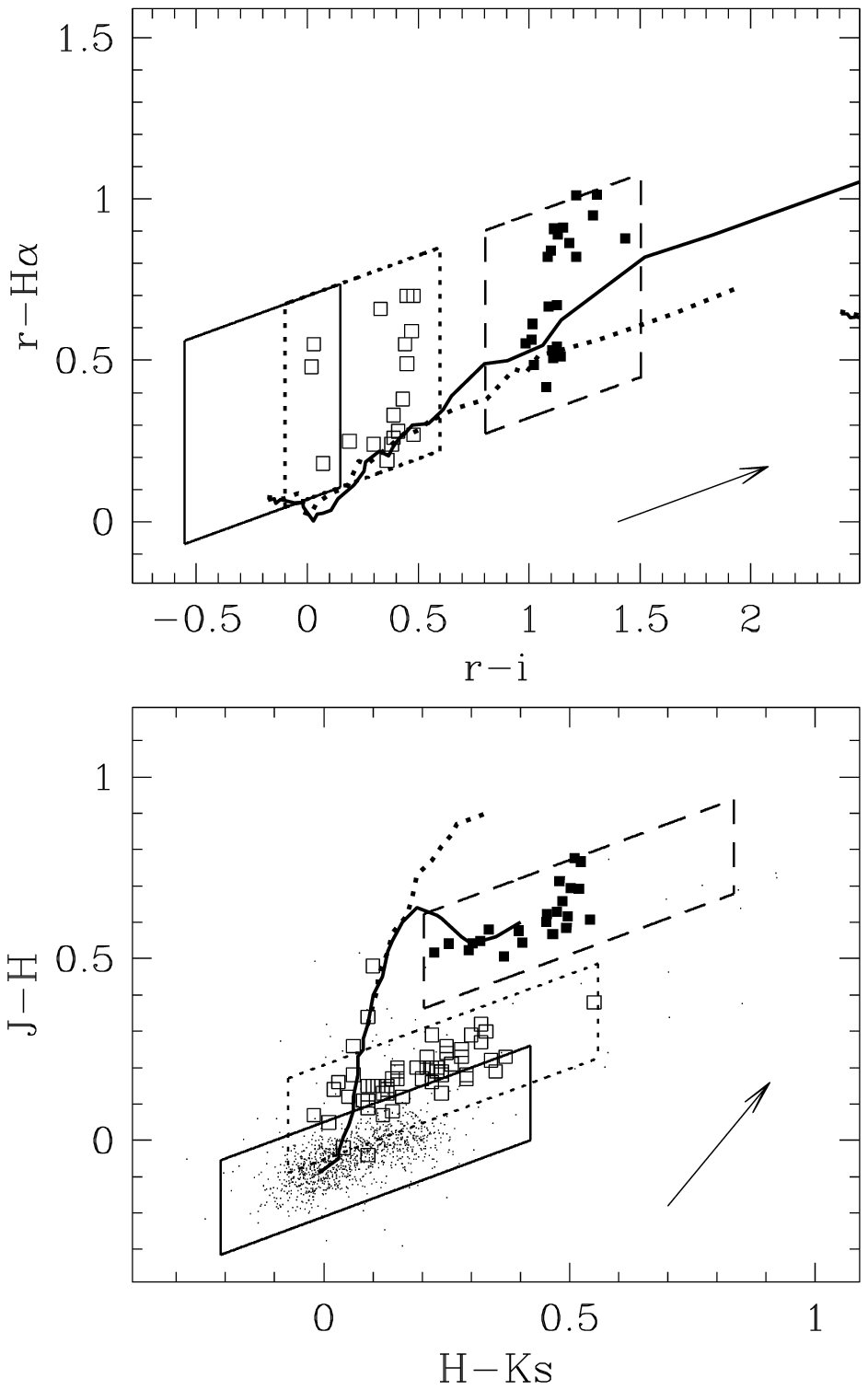}
\includegraphics[width=6.5cm]{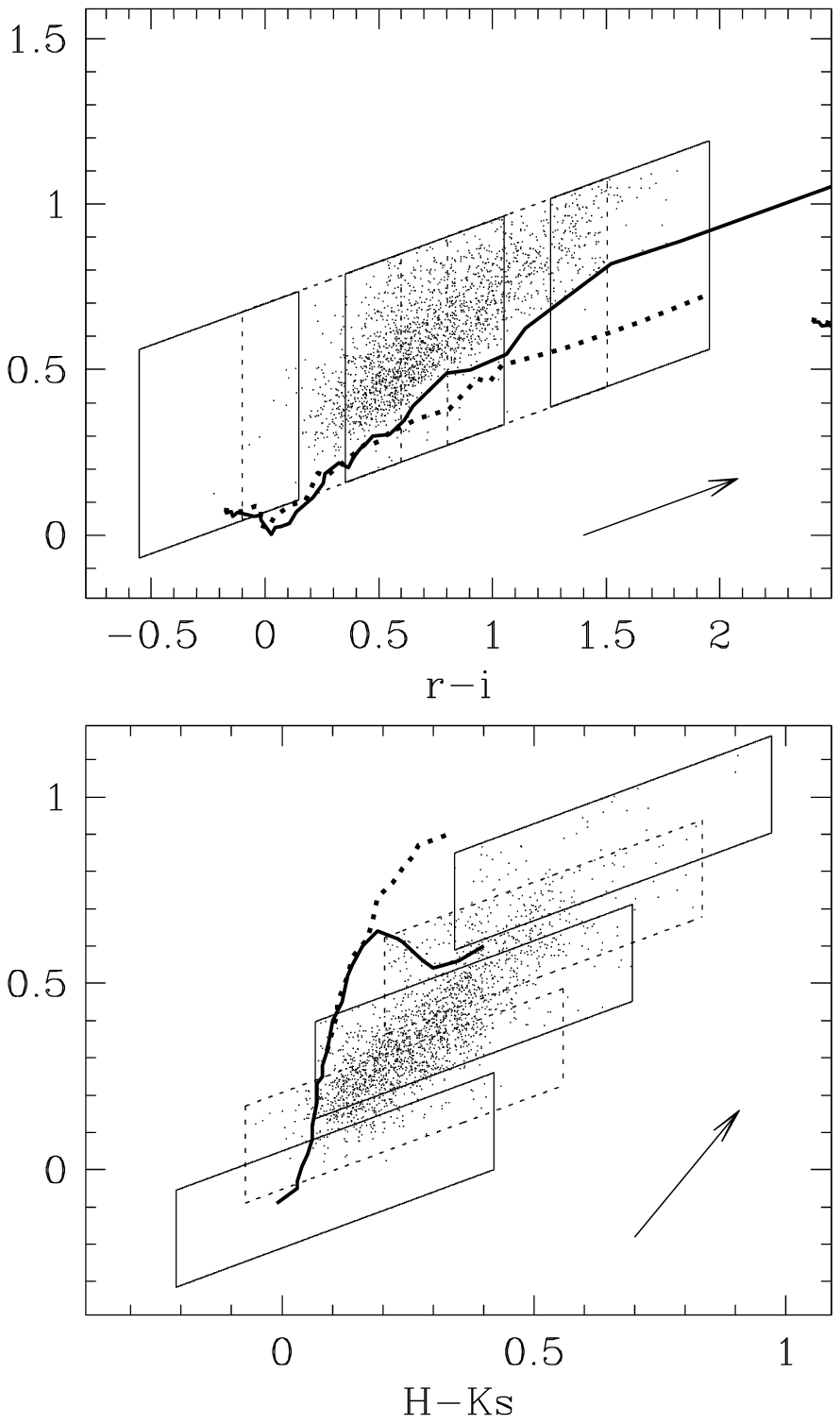}
\caption{{\bf Left:} location of known Be stars in the IPHAS and
2MASS diagrams.  The dots mark dereddened colours for the 1148 Be stars
from Zhang, Chen \& Yang (2005). Open squares are the reddened data
for the open clusters NGC~663, NGC~869 and NGC~884, and filled squares
similar data for NGC~7419.  The solid line is the adopted
zero-reddening selection box for Be stars.  {\bf Right:} the 2035
candidate Be stars in the list of 4338 \ha\ emitters by \cite{w08}.
The alternate solid/dotted boxes show the selection boxes (see text)
for $A_v$=0,2,4,6,8~mag, from bottom-left to top-right, respectively.
}
\label{F-Be}
\end{figure*}

In order to further investigate the similarities between
Fig.~\ref{F-2MASScc} and \ref{F-allwith}, we have studied in more
detail the location of Be stars in the colour-colour diagrams. Zhang,
Chen \& Yang (2005) provide the reddening of 1148 individual objects:
this allow us to identify the locus of unreddened Be stars in the
2MASS diagram, and to define a corresponding ``selection'' box. These
are indicated by the dots and the solid line, respectively, in the
lower-left panel of Fig.~\ref{F-Be}. When shifted to the reddening of
the clusters NGC~663, NGC~869 and NGC~884 ($A_V$$\sim$$2$ mag, dotted
line) and NGC~7419 ($A_V$$\sim$$6$ mag, dashed line), such a box
includes the corresponding data points for these clusters (empty and
filled squares, respectively).

Similar consistent results are obtained for the IPHAS colour-colour
diagram. There, if we define the locus of Be stars from the position
of the objects in the four open clusters considered, and correct for
the corresponding reddening, we can determine a zero-reddening
selection box for this class of objects, which is shown as a solid
line in the upper-left panel of Fig.~\ref{F-Be}.

We have then done the exercise of extracting candidate Be stars from
the 4338 IPHAS \ha\ emitters in \cite{w08}. An object is considered to
be a good candidate Be star if it falls inside the selection boxes in
{\it both} the IPHAS and the 2MASS diagrams.  The combination of the
selection boxes in the two diagrams, shifted so as to consider a range
of reddening values from $A_V$$=$$0$ to $A_V$$=$$8$~mag (represented by the
sequence of boxes in the right panels of Fig.~\ref{F-Be}), results in
a total of 2035 candidate Be stars (also plotted in
Fig.~\ref{F-Be}). The majority of them would have reddening values
between 3 and 5 magnitudes in V. Objects at zero reddening are
virtually absent, presumably because such objects are in the main nearby 
and thus bright and saturated in IPHAS.
%There are also few candidates with $A_V>7-8$~mag, because highly
%reddened and consequently likely distant, and thus undetected by
%IPHAS.
Only for significant reddening values (i.e. at the largest \jh\
colours, cf. the lower-right panel of Fig.~\ref{F-Be} with
Fig.~\ref{F-2MASScc}), mixing of the Be stars candidates with
low-reddening T~Tauri stars is possible. An idea of the magnitude of
the contamination can be gained considering that only 4\%\ of the 450
T~Tauri stars in \cite{ds05}, including those with the lowest \ha\
equivalent widths, have \jh$\le0.55$. Adopting this figure, and the
worst possible (and unlikely) scenario that {\it all} sources with
\jh$>0.55$ in our complete list of 4338 objects are T~Tauri stars, it
would be concluded that only one hundred objects of this class would be 
expected with a lower \jh\ colour, at variance with the more than 1600 
Be~star candidates (80\%\ of our list) lying below this colour limit. 

A smaller amount of mixing is expected with other, less frequent 
classes of objects, like CVs, compact PNe, and symbiotic stars.
We conclude that {\it somewhere between 1500 and 2000 \ha\ emitters in
\cite{w08} are likely to be Be stars}, which would therefore be the
largest class of objects in this list of IPHAS \ha\ stars.

\begin{figure}[!ht]
\centering
\includegraphics[width=8.8cm]{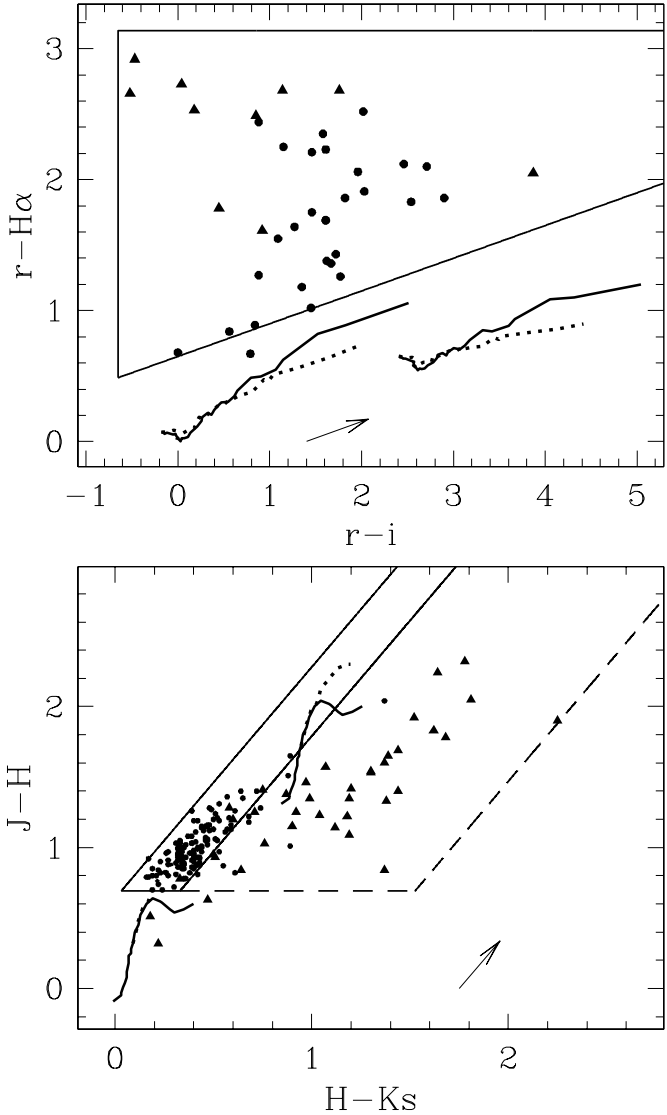}
\caption{{\bf Top}: the selection box for symbiotic stars in the IPHAS
colour-colour diagram, superimposed on the reference sample and the
locus of main-sequence and RGB stars (see Fig.~\ref{F-IPHAScc} for
details). {\bf Bottom}: the selection boxes for the S-type (continuous
line) and the D-type (dashed line) symbiotics in the 2MASS
diagram. Symbols as in Fig.~\ref{F-2MASScc}.}
\label{F-selbox}
\end{figure}

\subsection{Planetary nebulae}
\label{S-PNe}

In Sect.~\ref{S-IPHAScc}, we remarked on the overlap in the IPHAS
diagram between PNe and symbiotic stars, especially those of the
D-type. On the other hand, the 2MASS colours are generally different,
allowing a neat separation of the two classes of objects (with notable 
exceptions which might be related to misclassification).  
A search for compact PNe from the list of \cite{w08}, resulting in 
a few new candidates, and a confirmation of their nature by spectroscopic
observations, is presented elsewhere (Viironen et al. 2008).

\subsection{Symbiotic stars}
\label{S-selectsymbio}

In a similar way as done for Be stars, based on the properties of our
reference samples, we define selection boxes in the colour-colour
diagrams to be used for our search for symbiotic stars in the
Milky Way.

In the IPHAS diagram, the limited number of symbiotic stars of the
D-type in the reference sample, as well as their significant overlap
with the S-types (with possible differences, however, as outlined in
Sect.~\ref{S-IPHAScc}), prevent us from defining separate boxes for
the two classes. We prefer instead to use the same selection
criterion, leaving to the near-IR colours the task of separating the two
classes according to the original definition.  Our selection box for
symbiotic stars (Fig.~\ref{F-selbox}, upper panel) is then defined as
follows.  In the \ri\ axis, we allow the range of intrinsic colours
spanned by known objects as discussed at the end of
Sect.~\ref{S-IPHAScc}, and extend it to the right side of the graph to
allow for the significant extinction that is expected when observing
through the Galactic Plane.

In the vertical direction, we define a lower limit which is an
inclined line, parallel to the reddening vector (Sect.~\ref{S-IPHAScc}), 
and defined by the formula $(r-H\alpha) \ge 0.25 \cdot (r-i) + 0.65$,
which includes all known symbiotic stars in the samples of
Sect.~\ref{S-IPHAScc} except for the yellow symbiotic BD-213873. This
limit roughly corresponds to \ha\ equivalent widths of 50~\AA\ (Drew
et al. 2005).
%AG~Dra, and LT~Del, the extragalactic carbon star symbiotic Draco C~1,
%and SS~324.
We decided to fix the lowest side of the selection box at this limit
in order to avoid significant mixing with Be stars (see
Sect.~\ref{S-bestars}). The upper limit of the selection box for
symbiotic stars is the \rha\ value for pure \ha\ emitters, namely
\rha$\sim$3.1.

In the 2MASS diagram, the large number of objects available allows us
to define separate boxes for the S and D-type symbiotic stars.  For
the S-types, we first set the lower limit for the \jh\ colour which
includes all known objects.  Then the left and right limits for the
\hk\ colours were chosen to run parallel to the reddening vector. They
are chosen to include the vast majority of S-type systems, except for nine
objects (7\%\ of the whole sample)
% 128 Stypes in total 
that are detached and on the right side -- these will, in any case, be 
included in the selection box for D-types. We have decided to keep the 
right limit for S-types (the most frequent class of symbiotics) as leftward 
as possible, in order to minimize mixing with T~Tauri stars (even if
at the expense of the D-types).  The selection box for S-type
symbiotic stars is indicated by the solid line in the lower panel of
Fig.~\ref{F-selbox}.  For the D-types, we use the same procedure, with
the further assumption of setting the same lower limit for the \jh\
colour as for the S-types, and the \hk\ left limit so as to have
contiguous boxes with no gap or overlap.  The selection box for the
D-type symbiotics is indicated by a dashed line in Fig.~\ref{F-selbox}.

In this way, the three yellow symbiotics AS~201, StHA~190, and
V471~Per, fall out of both boxes in the 2MASS diagram.  But they
are located in the very populated region of the diagram where not only
main sequence stars are found, but also other \ha\ emitters like Be
stars and cataclysmic variables.  With so much mixing, a search for
this kind of object becomes very difficult.  

\begin{figure*}[!ht]
\centering
\sidecaption
\includegraphics[width=6.5cm]{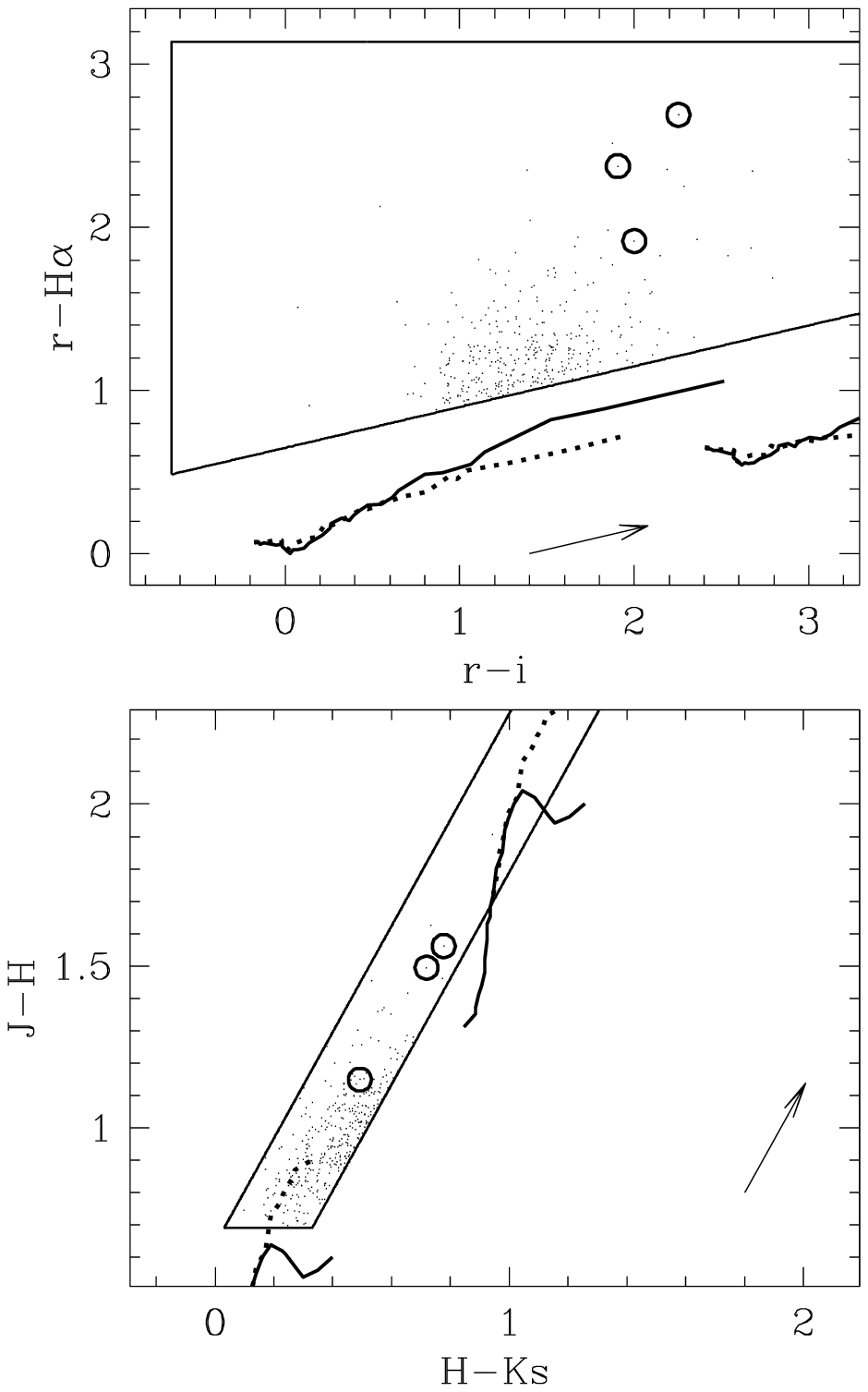}
\includegraphics[width=6.5cm]{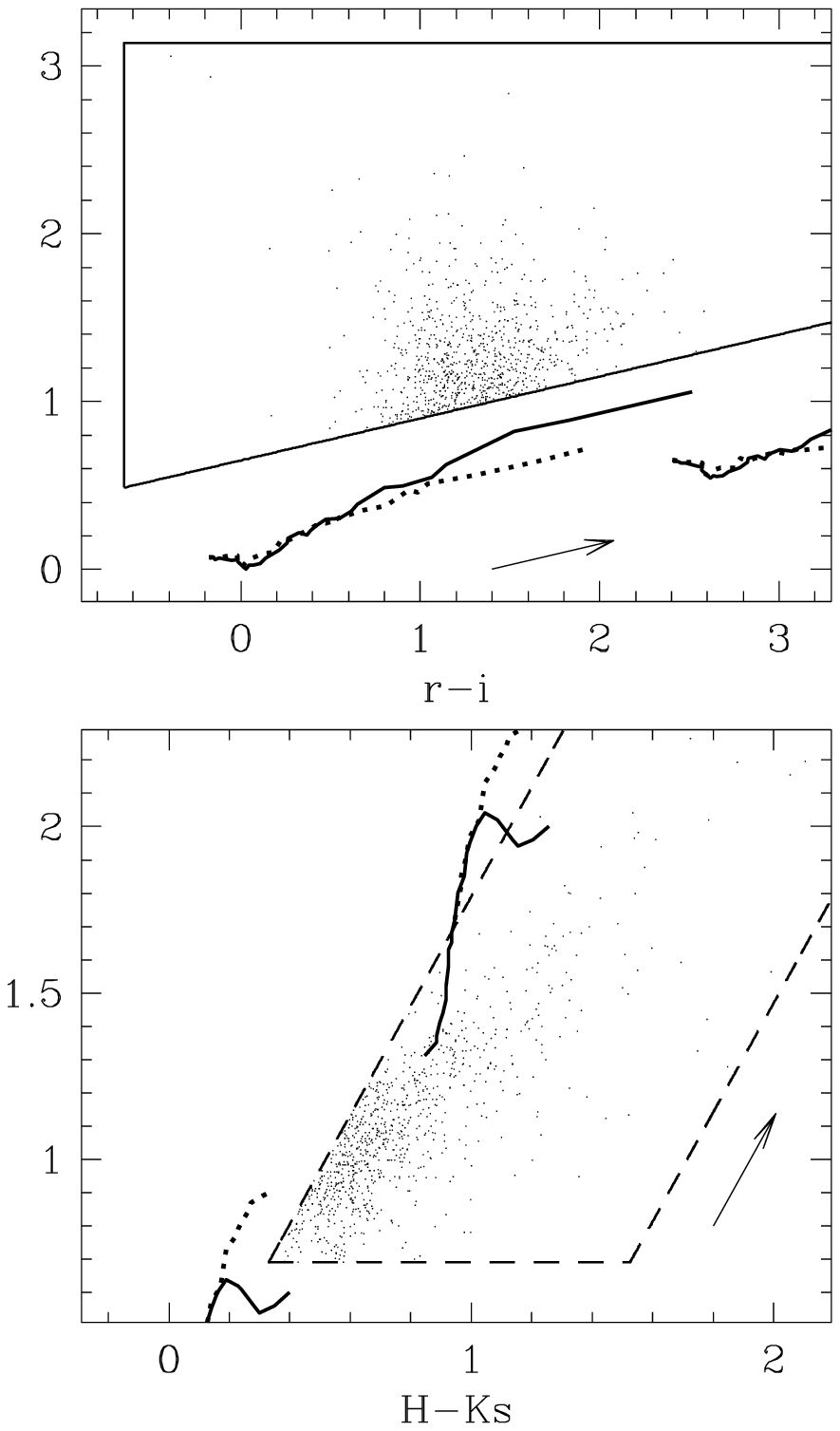}
\caption{{\bf Left:} The 337 objects with IPHAS and 2MASS colours
fulfilling our selection criteria for S-type symbiotic stars. The
locations of the three new symbiotic stars discovered by IPHAS, for
which we present spectroscopic confirmation in Sect.~\ref{S-spectra},
are indicated by the circles.  {\bf Right:} the 846 candidate
D-type symbiotic stars.}
\label{F-sselected}
\end{figure*}

When our selection criteria are applied to the list of \cite{w08}, 337
sources fall in both the IPHAS and 2MASS selection boxes for the
S-types (Fig.~\ref{F-sselected}, left), and 846 fulfill the criteria
to be considered D-type candidates (Fig.~\ref{F-sselected}, right).
The complete list of all these candidates is reported in
Tab.~\ref{T-selected}, which is only available electronically.  In the
2MASS diagram, a good number of objects fall in the lower right part
of the selection boxes for S-type symbiotic stars, while most of the
objects in the D-type box fall in its left-bottom side. These are the
regions where T~Tauri stars are also expected to be found, indicating
that significant mixing of the two classes of objects is likely to be
present in the candidate list (Tab.~\ref{T-selected}).

Our selection method recovers the three known symbiotic stars included
in the list of \cite{w08}, as well as two suspected ones. Out of the
29 objects listed by \cite{w08} as known young stellar objects, 13 are
included in the list of D-type candidate symbiotics, and 3 in the list of
S-type candidates (note that none of them is classified as a Be star).
This confirms that the mixing with young stellar objects is more
severe for the D-types. 
%A corresponding comment is included in Tab.~\ref{T-selected}.  
Among the 9 known or newly discovered PNe
included in the list of \cite{w08}, only one enters the list of D-type
candidates. Its nature will be discussed in Viironen et al. (2008).

% DATA in YSO_in_SDcand.lis and symbiopne_dist.dat

%Three are in Witham list of emitters:\\ 
%DQ~Ser        \\
%V1413 Aql     19 03 46.86 +16 26 17.0   IN WHITHAM LIST!\\
%PN Ap 3-1     19 10 36.126 +02 49 28.69 IN WHITHAM LIST!\\

Note that if the lowest edge of the IPHAS selection box is lowered so
as to include all known symbiotic stars in our samples, then another
152 candidates of S-type, plus 330 of D-type, would be added.  At the
same time, the confusion with other classes of object would also
increase.  In this respect, we stress that our aim is not to define
absolute limits for the colours of symbiotic stars, but only to
provide a practical way of selecting new candidates from the IPHAS
survey, guided by the best reference samples that we were able to
build from the literature, IPHAS itself, and from additional
observations obtained in backup time during our observational
campaign. As demonstrated below, the proposed selection method seems
indeed to be promising. 
%In addition, a continuous improvement of the selection criteria is
%expected as our spectroscopic follow up proceeds.

\subsection{Clustering as an additional criterion to separate  
young stars from symbiotics}
\label{S-clumping}

Observations targeting a narrow band of the Galactic Plane 
%implies 
must include a large number of star-forming regions, young clusters
and associations. There, numerous objects, especially Be and T~Tauri
stars, are \ha\ emitters. We have shown that while Be stars can be
separated from symbiotic stars because of their smaller \rha\ and \jh\
colours, T~Tauri stars overlap with symbiotics in both the IPHAS and
2MASS diagrams. Therefore, they must be regarded as the most serious
``contaminants'' in our search for symbiotic stars using IPHAS and
2MASS data.

Our strategy for tackling the problem is to consider the spatial 
distribution of the \ha\ emitters selected by IPHAS. T Tauri stars are
expected mainly to be concentrated in young clusters. On the other hand, 
symbiotic stars, which belong to an older Galactic stellar population
(bulge/thick-disc, Munari \& Renzini 1992), should be more isolated
\ha\ sources (except for projection effects through the Galactic
Plane).

\begin{figure}[!hb]
\centering
\includegraphics[width=8.8cm]{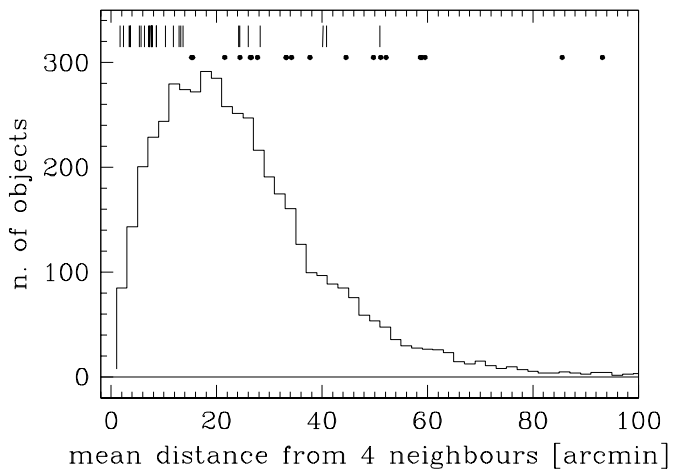}
\caption{Histogram of the mean distance to the four most nearby
objects for each of the 4853 \ha\ emitters by \cite{w08}. 29 young
stellar objects are indicated by the short lines at the top of the
diagram, while 20 PNe and symbiotic stars are indicated by the dots.}
\label{F-clump}
\end{figure}

A first step in exploring this issue was taken by measuring the degree
of clustering of the objects in the list of \cite{w08}. For each of
the 4853 \ha\ emitters, the mean distance to the $n$ most nearby
objects (with $n$ from 1 to 12) was computed. We show in
Fig.~\ref{F-clump} the distribution of the distances for $n=4$, that
we find to be a good compromise between having a sufficient number of
neighbours to detect the existence of a group, while avoiding the
limited statistics that the requirement of a large number of
neighbours might suffer from, given the size of the global sample.

The 29 objects listed by \cite{w08} as young stellar objects, as
well as 20 known, suspected or confirmed new PNe (Viironen et
al. 2008) and symbiotic stars (to be presented in paper II) were then
considered. They confirm the expected trend: young stars (the vertical
short lines at the top of Fig.~\ref{F-clump}) tend to occur with
smaller angular separations than do PNe and symbiotic stars (the filled
circles). Therefore, this additional parameter seems to be a useful
one to distinguish, at least in a statistical sense, young stellar
objects from stars belonging to older Galactic populations like
symbiotic binaries.  We expect that the significance of this parameter
will improve further as the sample of \ha\ emitters detected by IPHAS
grows in number; this would allow a better definition of clustering in
the area observed by IPHAS. For the time being, we add the mean distance 
to the four nearest neighbours as a further datum in the list of
candidate symbiotic stars in Tab.~\ref{T-selected}; the weight to be
given to this parameter is left to the discretion of the user. In any
case, we note that about 50\%\ of both the S-type and D-type candidates 
in Tab.~\ref{T-selected} have a mean distance to their four
nearest neighbours that exceeds 15~arcmin: these candidates are our 
preferred initial targets for spectroscopic follow-up.

\begin{figure*}[!ht]
\centering
\sidecaption
\includegraphics[width=15.4cm]{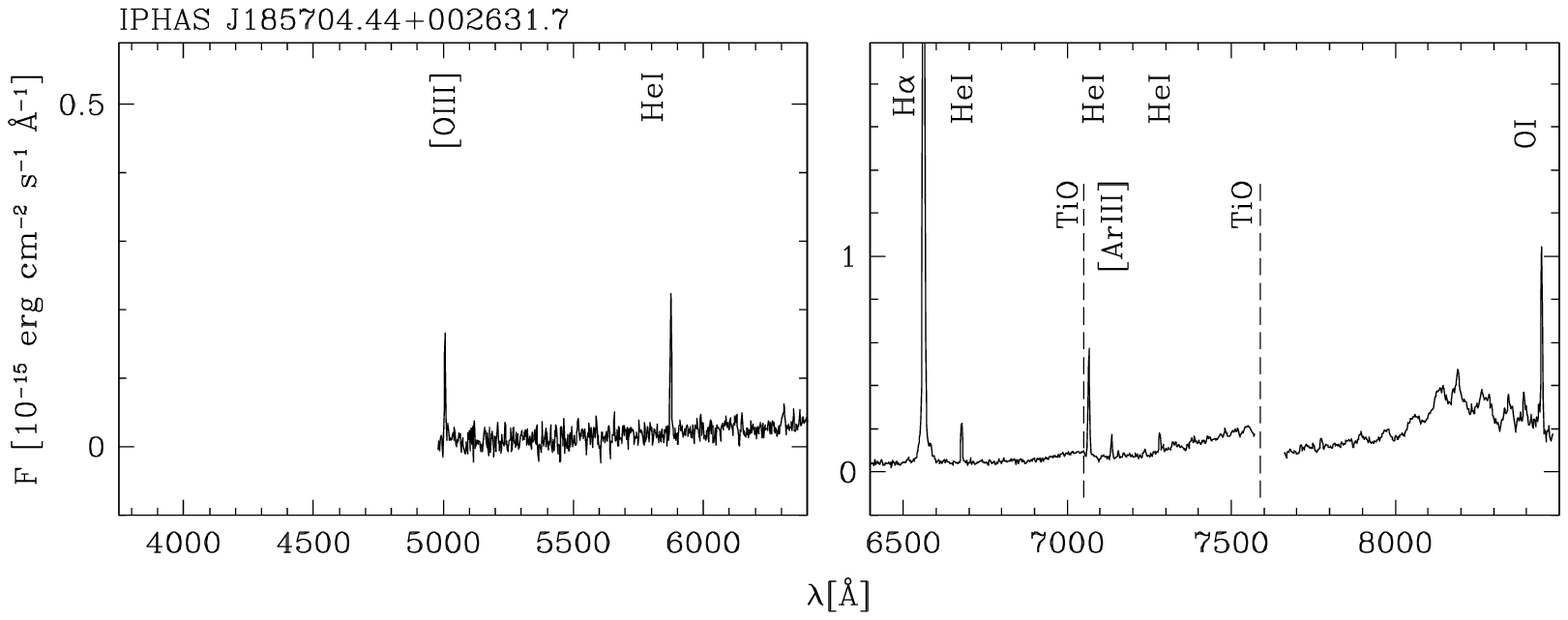}
\includegraphics[width=15.4cm]{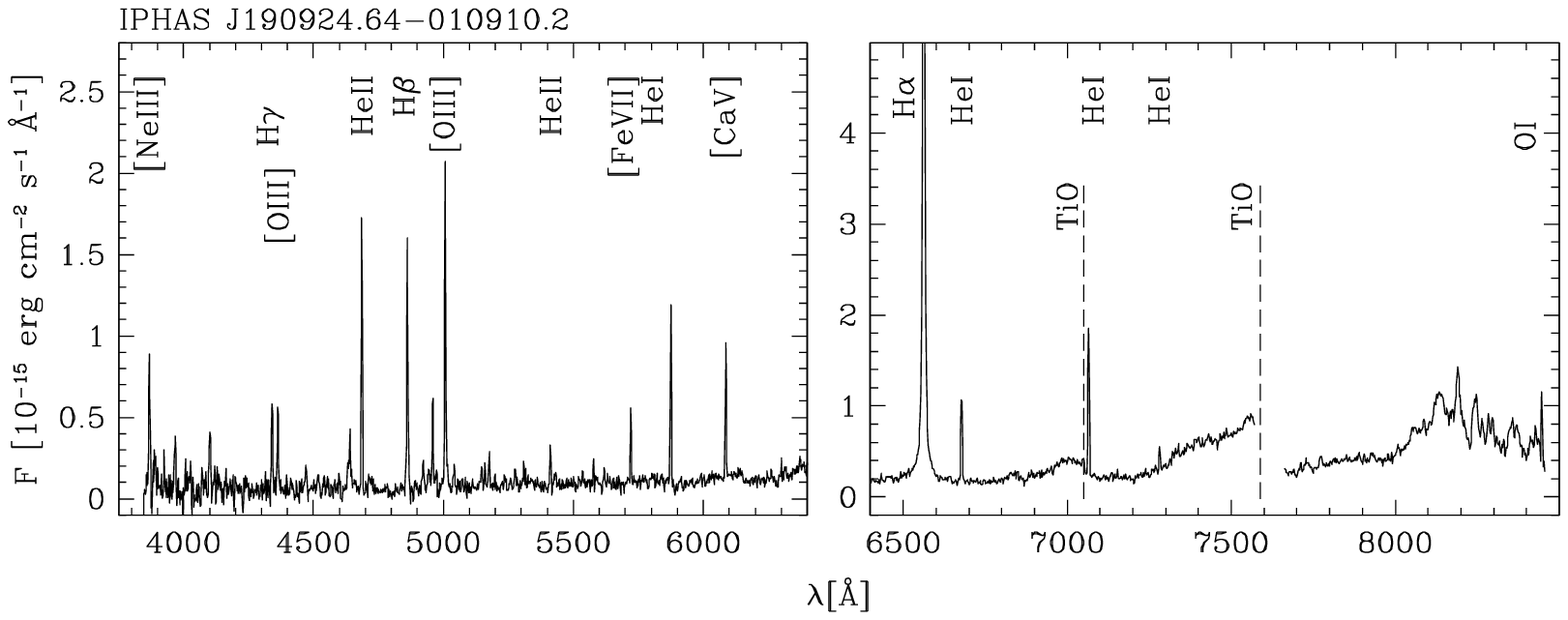}
\includegraphics[width=15.4cm]{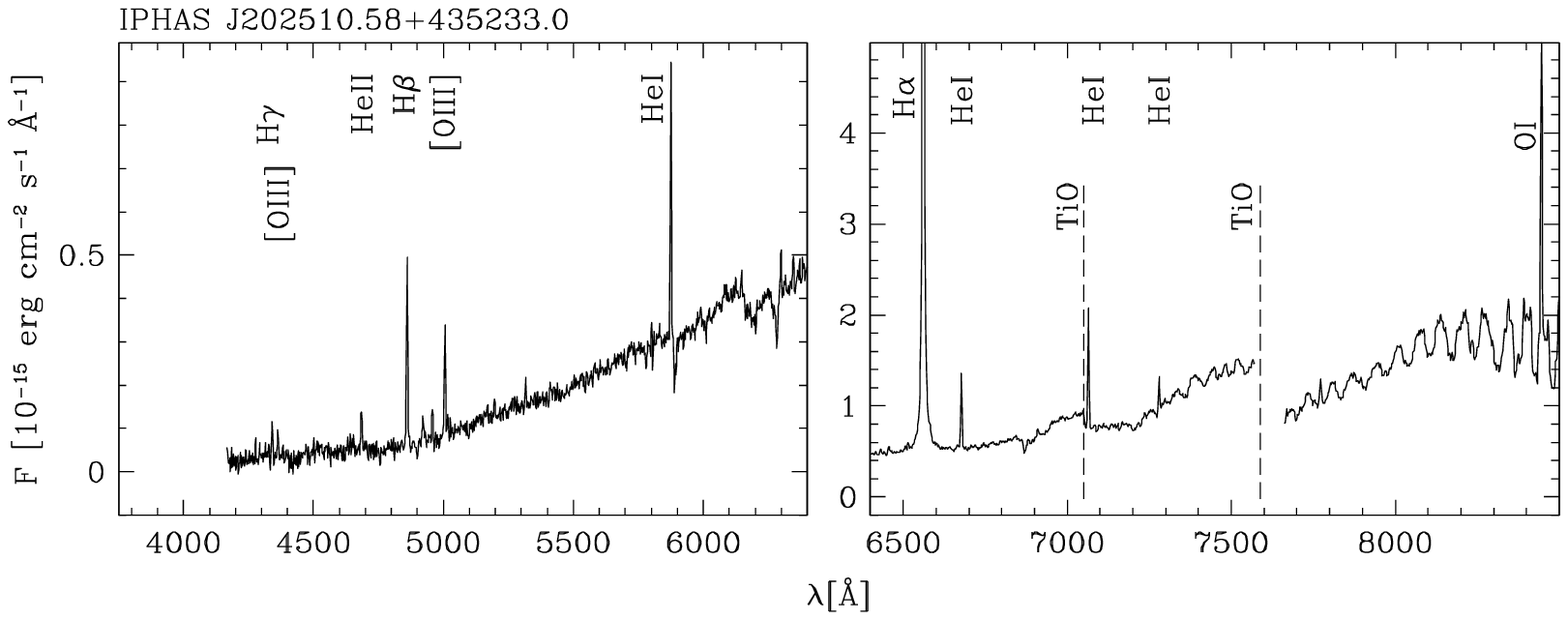}
\caption{Spectra of the first three new symbiotic stars discovered
by IPHAS. The region around 7600~\AA\ with the strong oxygen atmospheric
absorption band is not plotted. The main low and high excitation
emission lines and the heads of the TiO bands, which demonstrate the
symbiotic nature of the objects, are labeled. The oscillations in the
very red part of the spectra are due to fringing in the detector.}
\label{F-spectra}
\end{figure*}

\section {Spectroscopic confirmation: the first symbiotic 
stars discovered by IPHAS}
\label{S-spectra}

As an illustration of the potential of IPHAS and the selection method
proposed in the previous sections, we present here the spectra of the
first three new symbiotic stars discovered by IPHAS. Data for more
stars and a deeper analysis of the individual objects will be
presented in paper II. 
%In that paper, as our spectroscopic follow-up progresses and the
%sample of observed symbiotic candidates increases, the amount of
%mixing with other classes of objects as a function of the position in
%the colour-colour diagrams will be also discussed.

In the following, we use both the standard nomenclature for IPHAS
point sources (``IPHAS J'' followed by the J2000 coordinates, see Drew
et al. 2005), as well as the name ``IPHAS-Sy $nn$'' to indicate the
$nn$th symbiotic star discovered by the survey.

The INT spectra are shown in Fig.~\ref{F-spectra}, and the location of
the objects in the IPHAS and 2MASS diagrams is indicated by open
circles in Fig.~\ref{F-sselected}. The three objects have large \rha\
colours and fall in the selection box for S-type symbiotics. In the
latter, two of them have colours characteristic of strongly reddened
red giants, 
%($A_V$$>$$5$~mag), 
as confirmed by the INT spectra. 

\subsection{IPHAS J185704.44+002631.7 (IPHAS-Sy~1)}

The IPHAS and 2MASS magnitudes of IPHAS-Sy~1 are: $r$=18.28,
$i$=16.03, \ha=15.59, $J$=10.42, $H$=8.86, and $K_S$=8.09. No objects
are listed in the SIMBAD database, within a radius of 2 arcmin about the 
source coordinates, and the nearest known radio HII detection is over 30 
arcmin away.  There is no evidence to link this object to a known 
star-forming region.

The \ha\ equivalent width is $800$~\AA, which corresponds to
\rha=2.69.  This is the faintest objects of the three in the optical
(but not in the near-IR), and the one with the highest \jh\
colour. This can be mainly ascribed to reddening, as shown by the
steep decrease of the flux below 5000~\AA\ (no emission is detected
blueward of the [OIII]5007~\AA\ line) and the large $r-K_S$ colour
amounting to 10.2 magnitudes.  The symbiotic nature of IPHAS-Sy~1 is
indicated by the simultaneous presence of high excitation emission
lines of [OIII] and [ArIII], and the TiO absorption bands of a M-type
star (Fig.~\ref{F-spectra}, upper panel).

\subsection{IPHAS J190924.64-010910.2 (IPHAS-Sy~2)}

IPHAS-Sy~2 has the following magnitudes: $r$=17.16, $i$=15.26,
\ha=14.79, $J$=11.44, $H$=10.29, and $K_S$=9.80.  Its \ha\ equivalent
width is $650$~\AA\ (\rha=2.38).

In SIMBAD, it is indicated as a possible PN (K~4-17), but the spectrum
described by \cite{SA87} does not allow its real nature as a
symbiotic star to be identified.  This is instead revealed by our INT 
spectroscopy, which captures a rich set of high excitation emission lines 
([NeIII], [OIII], HeII, and even [FeVII]), accompanied by the red continuum 
of an M4 giant (Fig.~\ref{F-spectra}, middle panel). No star formation 
regions are catalogued in the vicinity of IPHAS-Sy~2.

\subsection{IPHAS J202510.58+435233.0 (IPHAS-Sy~3)}
\label{S-IPHASJ202510.58+435233.0}

Its magnitudes are: $r$=15.34, $i$=13.34, \ha=13.42, $J$=9.94,
$H$=8.45, and $K_S$=7.73.  The \ha\ equivalent width is
$250$~\AA\ (\rha=1.92).
% The spectral type of the cool component is M0.

IPHAS-Sy~3 is listed in SIMBAD as a misclassified planetary nebula
(PN~K~3-59), but the spectrum of \cite{SFO87} only reveals weak \ha\
line emission superimposed on a very red continuum. Our observations show
the composite spectrum defining a symbiotic star, with the
simultaneous presence of the continuum of an early M star and high
excitation emission lines.

In projection, the star is located in the Cygnus~X complex, a little over
3 arcmin from the HII region DWB~137 (\cite{dwb69}).  The vicinity of young 
stars might cast some doubt on the nature of IPHAS-Sy~3 as a symbiotic star.
Even if it shows some of the typical spectral signatures of these interacting
binaries, such as the TiO bands together with highly ionized HeII and
[OIII] in emission, some extreme examples of T Tauri stars also present them 
(see e.g. \cite{b01}).  

   However, there are several arguments pointing toward IPHAS-Sy~3 as
most likely being a symbiotic star. First, its spectrum lacks evidence
of other features typical of T~Tauri stars: e.g. the strongest
resonance Li doublet at 6708~\AA\ in absorption; any of FeII or NaI 
lines in emission; apparent broadening of the HeI emission line
profiles.  Second, T~Tauri stars, being closer to the main sequence
(MS), have radii just 2-3 times larger than their MS descendants,
whereas symbiotic stars must contain a red giant with radii two orders
of magnitudes larger than MS values.  Playing devil's advocate, we can
see if a typical T Tauri star radius works for IPHAS~Sy~3 if placed at
a plausible distance to be in a young Cygnus-X cluster.  The distance
to the young cluster containing the Herbig Be star V1685 Cyg, some 3
degrees away on the sky, is given as 980~pc (\cite{d01}), with the
other nearby HII regions in Cygnus~X being mostly at 1--2~kpc
(\cite{s93}).  For a distance of $\sim$1~kpc, an M0 spectral type
(compatible with the strength of the TiO and VO absorption bands), a
reddening E(B-V)$\sim$2--2.5 (estimated from HI and HeI emission line
ratios, the $J$$-$$K_S$ colour, and a fit of the optical and near-IR
spectral energy distribution), we apply the Barnes-Evans relation
(e.g.  \cite{b99}) to derive a radius $\sim$30~$R_\odot$.  This is too
large for a T~Tauri star, but consistent with the presence of a red
giant.  The estimated radius increases if the distance is larger than
1~kpc.
%Also, the K-band absolute magnitude of IPHAS-Sy~3 would be consistent
%with that of an M0III giant, and not of a lower luminosity object for
%a distance $\ge$1.5~kpc.

\section{Summary and  perspectives}

Searching for a relatively rare and old family of stars in a
narrow band in the Galactic plane, where young stellar objects
dominate the population of \ha\ emitters, might seem a 
hard challenge. Our motivation to embark in such an enterprise is the
poor knowledge of the total population of symbiotic stars in the
Galaxy:
%a basic figure to discuss the relevance of these systems to certain
%important astrophyis like thei long-standing problem of the origin of
%type Ia supernovae.
the IPHAS survey gives us the opportunity to tackle this problem by
performing, for the first time, a complete search of these objects in
a magnitude limited volume.  

As symbiotic systems contain a cool giant star, the near-IR data from
the 2MASS survey have been added to the IPHAS photometry.  Then, a
discussion of the location of the different classes of \ha\ emitters
in the combined IPHAS and 2MASS colour-colour diagrams has been
presented. This allows us to define selection criteria for symbiotic
stars which separate them
%, on the basis of their IPHAS and 2MASS colours, 
from the vast majority of normal and \ha\ emitting stars.

The only exception are T~Tauri stars. They overlap with symbiotic
stars in the IPHAS diagram. In the 2MASS one, they form a sequence
very close to (and partially overlapping with) the main subclass of
symbiotic stars (S-types), and fully overlap with the other, less
frequent group of symbiotics (the D-types). In fact, even when a
spectrum is available (see Sect.~\ref{S-IPHASJ202510.58+435233.0}), it
is sometimes difficult to distinguish a symbiotic star from those
T~Tauri stars which display high excitation emission lines.  As an
additional criterion to separate symbiotic from T~Tauri candidates in
our survey, we have considered the spatial distribution of the \ha\
emitters detected by IPHAS. T~Tauri stars are expected to be found in
groups in star forming regions.  In contrast, symbiotic stars
should be more isolated objects. With this aim, we have introduced a
``clustering'' parameter which is the mean distance to the four most
nearby \ha\ emitters in the IPHAS list that we have considered (Witham
et al. 2008).

We expect that all these criteria for the search for symbiotic stars
within IPHAS will improve as soon as additional information is
gathered from the survey itself. A better fix on the number of
\ha\ emission line stars above specified, more exacting equivalent
width thresholds, might improve our estimates of the clustering
parameter -- or further follow-up spectroscopy of the selected candidates
can improve our understanding of the IPHAS and near-infrared
colour-colour planes.
% Janet: HAVE REPLACED '2MASS' BY 'NEAR-INFRARED' TO ALLOW FOR THE
%ADVENT OF DEEPER UKIDSS DATA.

With this second aim in mind, we have started an intensive spectroscopic
campaign. With it, we hope that the number of symbiotic stars in the
IPHAS area, so far limited to only 11 objects listed in the catalogue
of Belczy\'nski et al. (2000), will significantly increase. The
confirmation in this paper of three new systems out of a small sample
of candidates observed in the first nights of the spectroscopic
follow-up, seems to be a good start in this direction.

\begin{table}[b]
\caption{1183 candidate symbiotic stars extracted from the list of
\cite{w08} using the photometric constraints defined in
Sect.~\ref{S-selectsymbio}.  Columns contain: the coordinates, the
IPHAS and 2MASS magnitudes and their errors, the mean distance $d_4$
(in arcmin) from four neighbours (see Sect.~\ref{S-clumping}).
Tab.~1a contains the candidate S-type symbiotics, and Tab.~1b the
candidate D-types.  The Table is available only in electronic form via
the CDS.}  \centering
%\begin{tabular}{ccccc|cr}
%\hline\hline\noalign{\smallskip}
%% Name & ...
%\end{tabular}
\label{T-selected}
\end{table}

\begin{acknowledgements}
We are grateful to many of our collaborators in the IPHAS project, for
continuous discussion about the properties of the variety of objects that
are involved in the analysis of the survey data. RLMC, ERRF, AM, and
KV acknowledge funding from the Spanish AYA2002-0883 grant, and JM funding 
from Polish KBN 1PO3D 017 27 grant.

\end{acknowledgements}

\end{document}